\documentclass[conferemce]{IEEEtran}
\usepackage{graphicx}
\usepackage{epstopdf}
\usepackage{amsthm}
\usepackage{epsfig}
\usepackage{latexsym}
\usepackage{amsfonts}
\usepackage{here}
\usepackage{rawfonts}
\usepackage[latin1]{inputenc}
\usepackage[T1]{fontenc}
\usepackage{calc}
\usepackage{capitalgreekitalic}
\usepackage{url}
\usepackage{enumerate}
\usepackage{color}
\usepackage[tbtags]{amsmath}
\usepackage{amssymb}
\usepackage{upref}
\usepackage{epic,eepic}
\usepackage{times}
\usepackage{dsfont}
\usepackage{comment}
\usepackage{cite}
\usepackage{bbm}
\usepackage{bm}
\usepackage{amsmath}
\usepackage{booktabs}
\usepackage{makecell}
\usepackage{csquotes}
\usepackage{multirow}
\usepackage{subfigure}

\usepackage{multirow}
\usepackage{array}
\usepackage{subcaption}
\usepackage{float}

\allowdisplaybreaks[4]

\usepackage{dsfont}

\newlength{\aligntop}
\newlength{\alignbot}
 

\IEEEoverridecommandlockouts

\usepackage{algorithm}
\usepackage{algorithmic}

\makeatletter
\newcommand\semihuge{\@setfontsize\semihuge{19.3}{25}}
\makeatother

\makeatletter
\newcommand\semismall{\@setfontsize\semihuge{12.4}{15}}
\makeatother

\usepackage{subfigure}

\begin{document}
\title{\huge
Language-oriented Semantic Communication for Image Transmission with Fine-Tuned Diffusion Model
}
\author{\IEEEauthorblockN{
Xinfeng Wei,
Haonan Tong, 
Nuocheng Yang, 
and Changchuan Yin}\\
\vspace{0.3cm}
\IEEEauthorblockA{\small Beijing Laboratory of Advanced Information Network, Beijing University of Posts and Telecommunications, Beijing, China\\
\thanks{This work was supported in part by Beijing Natural Science Foundation under Grant L223027, the National Natural Science Foundation of China under Grants 62471056, 61629101 and 61671086, the 111 Project under Grant B17007, in part by BUPT Excellent Ph.D. Students Foundation under Grant CX2021114, and China Scholarship Council.}
}
Emails: 
\{xinfengwei, hntong, yangnuocheng, and ccyin\}@bupt.edu.cn

}
\maketitle

\pagestyle{empty}  
\thispagestyle{empty} 
\begin{abstract}
Ubiquitous image transmission in emerging applications brings huge overheads to limited wireless resources. Since that text has the characteristic of conveying a large amount of information with very little data, the transmission of the descriptive text of an image can reduce the amount of transmitted data. In this context, this paper develops a novel semantic communication framework based on a text-2-image generative model (Gen-SC). 
In particular, a transmitter converts the input image to textual modality data.
Then the text is transmitted through a noisy channel to the receiver.
The receiver then uses the received text to generate images. 
Additionally, to improve the robustness of text transmission over noisy channels, we designed a transformer-based text transmission codec model. Moreover, we obtained a personalized knowledge base by fine-tuning the diffusion model to meet the requirements of task-oriented transmission scenarios.
Simulation results show that the proposed framework can achieve high perceptual quality with reducing the transmitted data volume by up to 99\% and is robust to wireless channel noise in terms of portrait image transmission.

\end{abstract}


\IEEEpeerreviewmaketitle

\section{Introduction}
In the future, data traffic of merging services is expected to continue increasing, posing significant challenges to resource-limited wireless networks. Especially,  image transmission in extremely resource-constrained (i.e., spectrum and power) and harsh environments remains challenging. Therefore, transmitting images using fewer network resources has potential applications across various scenarios. 

 Semantic communication is an emerging research paradigm expected to enable efficient data transmission \cite{10405124,9685654, 10622730, 10118717}. In \cite{bourtsoulatze2019deep}, the deep joint source and channel coding  (DeepJSCC) schemes have been developed to effectively compress and transmit images by optimizing the joint coding scheme to adapt to wireless channels. DeepJSCC encodes images based on the semantic features of the data to be transmitted, maintaining the visual information while achieving efficient image transmission. 

However, in some cases (i.e., task-oriented communication), transmitting all features of an image is unnecessary. For instance, in the retail industry, a customer's facial expressions can reveal their preference for a product. In AR/VR applications, user expressions can enhance gaming or human-computer interaction experiences. In these scenarios, the image receiver's focus is not on the people's specific identity in the image but rather on the people's state, expressions, and other contextual information. Therefore, In order to reduce the transmission volume, information unrelated to the portrait state should be removed, and the remaining effective information can be compressed into text form to meet the needs of such communication scenarios.

Recently, generative artificial intelligence (GenAI) models have seen significant advancements, a large amount of innovative methods utilizing GenAI models have been introduced in the realm of semantic communication. \cite{nam2024language} proposed an image transmission method based on descriptive information contained in images. In this method, the transmitter sends only the descriptive text information extracted from the original image using an image-to-text (I2T) algorithm to the receiver, and then the receiver reconstructs the image based on the received information using an image generation model. Textual data has the characteristic of conveying a large amount of information with very little data. In noisy channels, even a single letter error in the decoded text at the receiving end can lead to a meaning vastly different from the original. Therefore, it is necessary to enhance the noise resistance of text transmission methods. Moreover, without specific control input, it is difficult for generative models to generate high-fidelity portraits or styles from text alone.

\begin{figure*}[htbp]
    \centering
    \includegraphics[width=12cm]{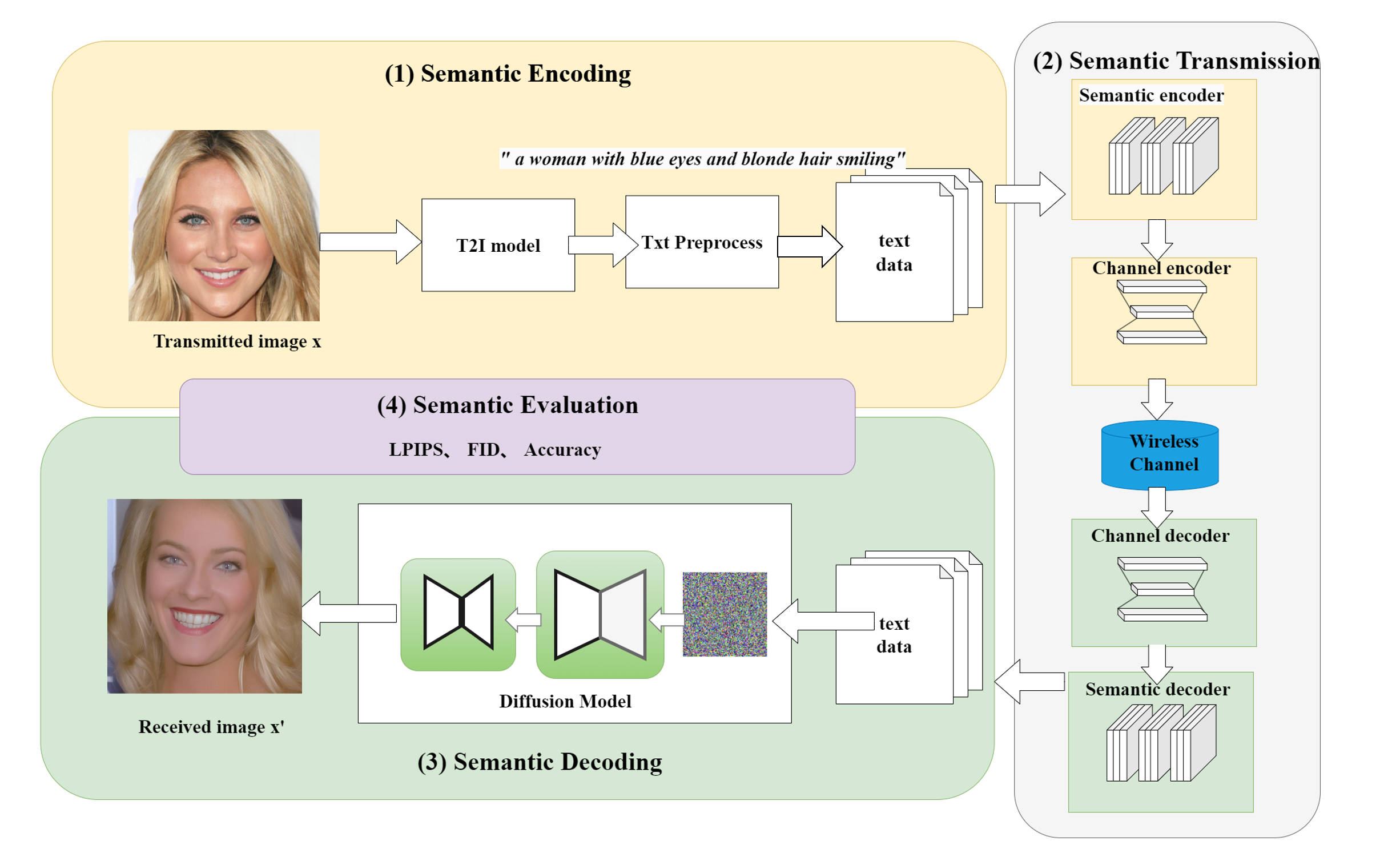}
    \caption[The framework of semantic communication for networks.]{\centering The framework of semantic communication for networks.}
    \label{fig1}
\end{figure*}

This paper aims to address the above issues, 
the main contributions are: 
\begin{itemize}
    \item We propose a novel text-to-image semantic communication system. In the system under consideration, the transmitter converts images to text, transmits the text using a deep learning-based end-to-end text communication method, and reconstructs the images at the receiver using a text-to-image generative model. 
    \item We designed transformer-based text transmission codec, enhancing the robustness of text transmission over noisy channels. 
    \item  We performed few-shot fine-tuning on the base diffusion model to generate high-fidelity portrait images so as to meet the needs of portrait image communication scenarios.
\end{itemize}
Simulation results demonstrate that in the task of portrait transmission, the proposed method achieves high perceptual similarity while effectively reducing the data volume and exhibiting robustness to noise.
The remainder of this paper is organized as follows. The system model is presented in Section \uppercase\expandafter{\romannumeral2}. We give a detailed description of the proposed semantic encoder and decoder model in section \uppercase\expandafter{\romannumeral3}. The simulation results are presented and analyzed in Section \uppercase\expandafter{\romannumeral4}. The conclusion is given in Section \uppercase\expandafter{\romannumeral5}.

\section{System Model And Problem Formulation}
  As shown in Fig. 1, our proposed framework consists of three main modules: semantic encoder through the img2txt model, semantic transmission involving the encoding and decoding of the text, and semantic decoder to reconstruct images based on received text.
\subsection{Semantic Encoder}
At the transmitter, the transmitter generates descriptive text data from the input image $\boldsymbol{v}$ using a pre-trained image-to-text~(Img2Txt) model. The text data retains semantic alignment with the image data which contains the intended object information of the source image. The text data is presented as a sentence $\boldsymbol{s}$ in a specific order and is given by: 
\begin{equation}
    \begin{aligned}
        \boldsymbol {s}=\operatorname{I2T}(\boldsymbol {v})=\left(\mathbf{s}_1, \mathbf{s}_2, \cdots, \mathbf{s}_{|\mathbf{S}|}\right),
    \end{aligned}
\end{equation}
where 
the function $\operatorname{I2T}(\cdot)$ represents an I2T encoder (i.e., bootstrapping language-image pre-training(BLIP)\cite{li2022blip}). Through modality conversion, the data volume can be significantly reduced, and data redundancy can be greatly minimized.  
\subsection{Semantic Transmission}
As shown in Fig.2, the sentence $\boldsymbol{s}=\left(s_1, s_2, \cdots, s_{|S|}\right)$ is fed into the transmission model, which consists of a transmitter and a receiver. The transmitter includes a semantic encoder and a channel encoder, both are implemented with neural networks. On the transmitting side, the semantic encoder extracts semantic information from $\boldsymbol{s} $  and maps to symbols $\boldsymbol{x}$, which are then transmitted over the physical channel by the channel encoder. Let the neural network parameters of the semantic encoder and the channel encoder be denoted as  $\beta $  and  $\alpha $, respectively. The encoded symbols$\boldsymbol{x}$  can be represented as:
\begin{equation}
\begin{split}
  \boldsymbol{x}=C_\alpha\left(S_\beta(\boldsymbol{s})\right),
\end{split}
\end{equation}
where $S_\beta(\cdot)$ represents the semantic encoder with parameter $\beta$ and $C_\alpha(\cdot) $ represents the channel encoder with parameter $\alpha$. The encoded symbols $\boldsymbol{x}$ are transmitted over the physical channel, assuming  $\boldsymbol{x}$ is normalized. The wireless channel is represented as $P_h(\boldsymbol{Y} \mid \boldsymbol{x})$, with $\boldsymbol{x}$ being the input and $\boldsymbol{y}$ being output. The transmission process through the wireless channel is given by 
\begin{equation}
 \boldsymbol{y} =\boldsymbol{h}\boldsymbol{x}+\boldsymbol{n},
\end{equation}
where $h$ indicates the channel gain, and $\boldsymbol{n} \sim \mathcal{N} \lvert(0, \sigma_n^2 \lvert \boldsymbol{I})$ is additive white Gaussian noise (AWGN) with variance $\sigma_n^2$ and $\boldsymbol{I}$ being identity matrix.
The receiver comprises a channel decoder and a semantic decoder to recover the transmitted symbols and decode the sentence. 
The decoded sequence is given by 
\begin{equation}
 \hat{\mathbf{s}}=S_\chi^{-1}\left(C_\delta^{-1}(\boldsymbol{y})\right).
\end{equation}
\subsection{Semantic Decoder}
We equip the receiver with a conditional image generative model, specifically utilizing Stable Diffusion\cite{rombach2022high} (SD), which has proven to be highly effective in generating images from textual descriptions. SD synthesizes an image starting from random noise and gradually refines it through a denoising process that is guided by a text prompt. This model consists of three main components: an encoder, which generates latent vectors into latent space from RGB image input; a denoiser, which performs the diffusion denoising process; and a decoder, which reconstructs the image from the latent vectors into RGB space. As depicted in Fig. 1, the diffusion model, under the guidance of the conditional text, iteratively reduce the noise by denoiser.

\begin{figure}[t]
\centering
\setlength{\belowcaptionskip}{-0.45cm}
\includegraphics[width=8.5cm]{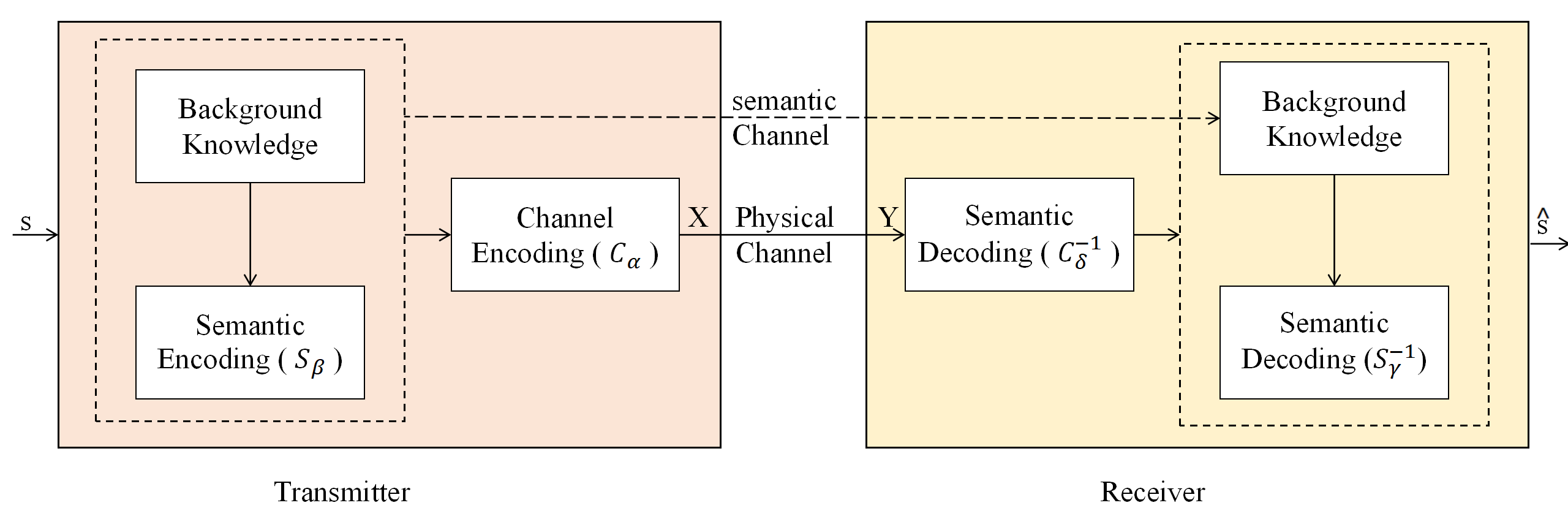}
\centering
\caption{The framework of end-to-end text transmission system.}
\label{fig2}
\end{figure}
\vspace{0.4cm}

\section{Networks architecture}

\subsection{Text Transmission model}
We use a transformer-based joint source-channel coding model, DeepSC \cite{xie2021deep}, as the text transmission codec. To improve the effectiveness of Deep-SC in noisy channels, we introduced the bidirectional and Auto-Regressive transformers model (BART) \cite{lewis2019bart}, resulting in BART-SC. The BART-SC is shown in Fig.3. The transmitter is composed of a semantic encoder and a channel encoder, the input sequence $\boldsymbol{s}$ is encoded by the semantic encoder, compressed by the channel encoder for transmission. The receiver consists of a channel decoder and a semantic decoder. The channel decoder decompress the received data, and finally semantic decoder reconstructs the original semantic information. 
Specifically, the core of the encoder section is a module consisting of three transformer layers, each including multi-head self-attention mechanisms and a feed-forward neural network\cite{vaswani2017attention}. The channel encoder compresses the output of the encoder into a low-dimensional representation using two fully connected neural networks. The channel decoder recovers the received low-dimensional representation to its original high-dimensional form using fully connected neural networks and performs normalization through a layerNorm layer. The final output is a probability distribution over the target vocabulary, generated through a fully connected layer.

The BART model is a transformer-based sequence-to-sequence pre-training model that learns to compress, generate, and decompress text by applying random masking and text generation to the original input sequences. BART employs a special masking method called "noise mask" to generate noise in the input and enable self-supervised training. Additionally, BART uses an autoregressive approach during training, where the decoder generates the output step by step to ensure semantic consistency between the generated output and the original input. Based on the above reasons, we introduce the BART module into the DeepSC.
The loss function is defined as:
\begin{equation}
    \begin{aligned}
        \mathcal{L}_{\text {total }}=\mathcal{L}_{\mathrm{CE}}(\mathbf{s}, \hat{\mathbf{s}} ; \boldsymbol{\alpha}, \boldsymbol{\beta}, \boldsymbol{\gamma}, \boldsymbol{\delta})+\lambda \mathcal{L}_{\mathrm{MI}}(\boldsymbol{x}, \boldsymbol{y} ; T, \boldsymbol{\alpha}, \boldsymbol{\beta}),
    \end{aligned}
\end{equation}
The first component is a cross-entropy loss function that measures the difference between $\hat{\mathbf{s}}$ and ${\mathbf{s}}$. 
\begin{equation}
    \begin{aligned}
        \begin{aligned}
        \begin{aligned} & \mathcal{L}_{\mathrm{CE}}(\mathbf{s}, \hat{\mathbf{s}} ; \boldsymbol{\alpha}, \boldsymbol{\beta}, \boldsymbol{\chi}, \boldsymbol{\delta})= \\ & -\sum_{l=1} q\left(w_l\right) \log \left(p\left(w_l\right)\right)+\left(1-q\left(w_l\right)\right) \log \left(1-p\left(w_l\right)\right)\end{aligned}
        \end{aligned},
    \end{aligned}
\end{equation}
For the $l\text{-th}$ word in the estimated sentence, denoted as $w_l$,  $q\left(w_l\right)$ represents its actual probability, while $p\left(w_l\right)$ represents the predicted probability of $w_l$ appearing in the sentence $\hat{\mathbf{s}}$.

\begin{figure}[t]
\centering
\setlength{\belowcaptionskip}{-0.45cm}
\includegraphics[width=8.5cm]{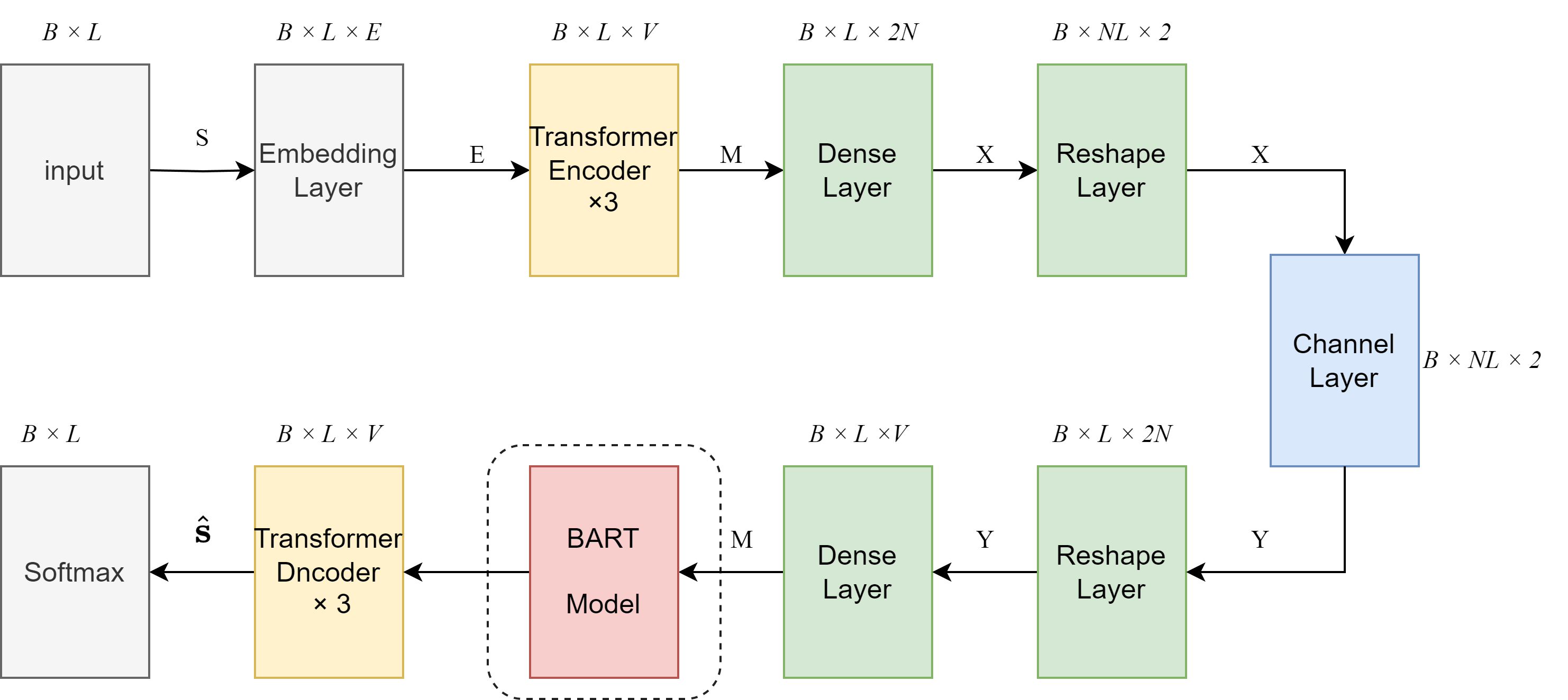}
\centering
\caption{The proposed neural network structure for end-to-end text transmission system.}
\label{fig2}
\end{figure}
\vspace{0.4cm}

The second component is a mutual information loss function that quantifies the amount of information shared between the transmitted symbol, which aims to maximize the data rate during transmitter training.
 \(I(X; Y)\) is the mutual information between the transmitted symbols \(x\) and the received symbols \(y\).
\begin{equation}
    \begin{aligned}
      \mathrm{I}(x, y)&=\int x \times y \times p(x, y) \log \frac{p(x, y)}{p(x) p(y)} d x d y
      \\ &=\mathbb{E}_{p(x, y)}\left[\log \frac{p(x, y)}{p(x) p(y)}\right],
    \end{aligned}
\end{equation}
We can optimize the encoder by maximizing the mutual information, which is expressed as:
\begin{equation}
    \begin{aligned}
       \mathcal{L}_{\mathrm{MI}}(\boldsymbol{x}, \boldsymbol{y} ; T)=\mathbb{E}_{p(x, y)}\left[f_T\right]-\log \left(\mathbb{E}_{p(x) p(y)}\left[e^{f_T}\right]\right),
    \end{aligned}
\end{equation}
where $f_T$ is composed by a neural network, in which the inputs are samples from $p(x, y)$, $p(x)$, and $p(y)$.

We implement a cross-training strategy to train BART-SC, we alternate training between the channel encoder/decoder and the semantic encoder/decoder models. Specifically, first, train the channel model and then freeze its parameters. Next, train the semantic model, freeze its parameters, and then retrain the channel model. This process can be repeated until both the semantic communication model and the channel model converge.

\subsection{Image generation with fine-tuned stable diffusion}
At the receiver, SD is utilized to generate the original image based on the received text. 
SD operates in a latent diffusion model, meaning it works within an autoencoder framework. Specifically, images are first encoded into a latent space by an encoder $\mathcal{E}$. The diffusion and reverse processes are applied in this latent space, and the resulting latent representations are then decoded back into image space by a decoder $\mathcal{D}$.
More precisely, in the diffusion process, noise is progressively added to the original latent tensor $x_0$ which is converted by autoencoder from input image. The model iteratively adds Gaussian noise to $x_0$:
\begin{equation}
\begin{aligned}
q\left(x_t \mid x_{t-1}\right)=\mathcal{N}\left(x_t ; \sqrt{1-\beta_t} x_{t-1}, \beta_t I\right), t=1, . ., T,
\end{aligned}
\end{equation}
where $q\left(x_t \mid x_{t-1}\right)$ is the conditional density of $x_{t}$ given $x_{t-1}$, and $\left\{\beta_t\right\}_{t=1}^T$ are hyperparameters. $T$ denotes the diffusion step. The reverse process is central to model training. The model learns to recover the clean latent tensor $x_0$ from a noisy latent tensor $x_t$:
\begin{equation}
\begin{aligned}
p_\theta\left(x_{t-1} \mid x_t\right)=\mathcal{N}\left(x_{t-1} ; \mu_\theta\left(x_t, t\right), \Sigma_\theta\left(x_t, t\right)\right),
\end{aligned}
\end{equation}
for $t = T,...,1$, which allows to generate a valid signal $x_0$ from the standard Gaussian noise step by step.  Finally, $x_0$ is decoded back to the RGB space by autoencoder's decoder to get the generated image.

\begin{figure}[t]
\centering
\setlength{\belowcaptionskip}{-0.cm}
\includegraphics[width=9cm]{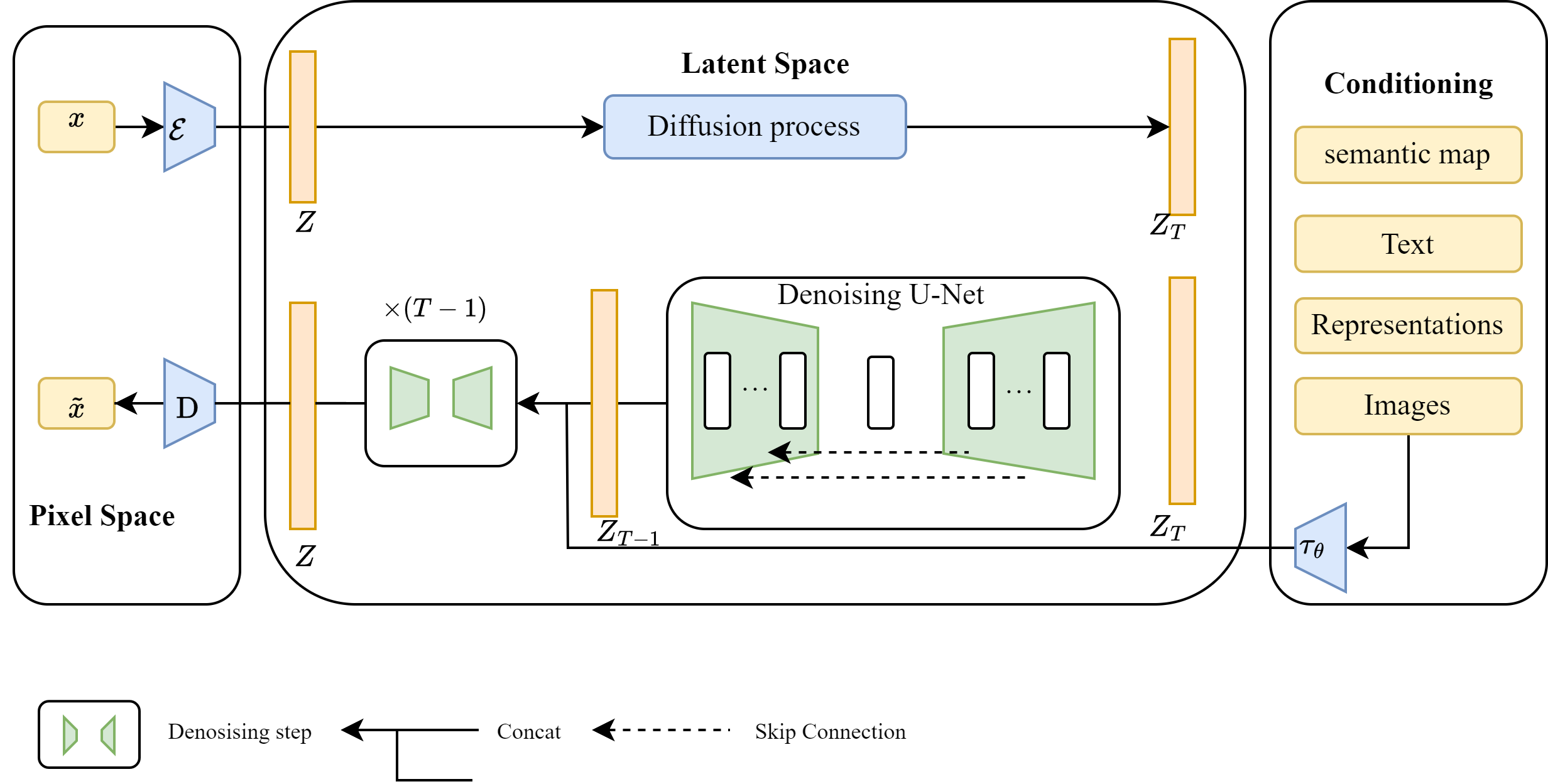}
\centering
\caption{The architecture of Stable Diffusion model.}
\label{fig3}
\end{figure}


\begin{figure}[t]
\centering
\setlength{\abovecaptionskip}{0.4cm}
\includegraphics[width=8cm]{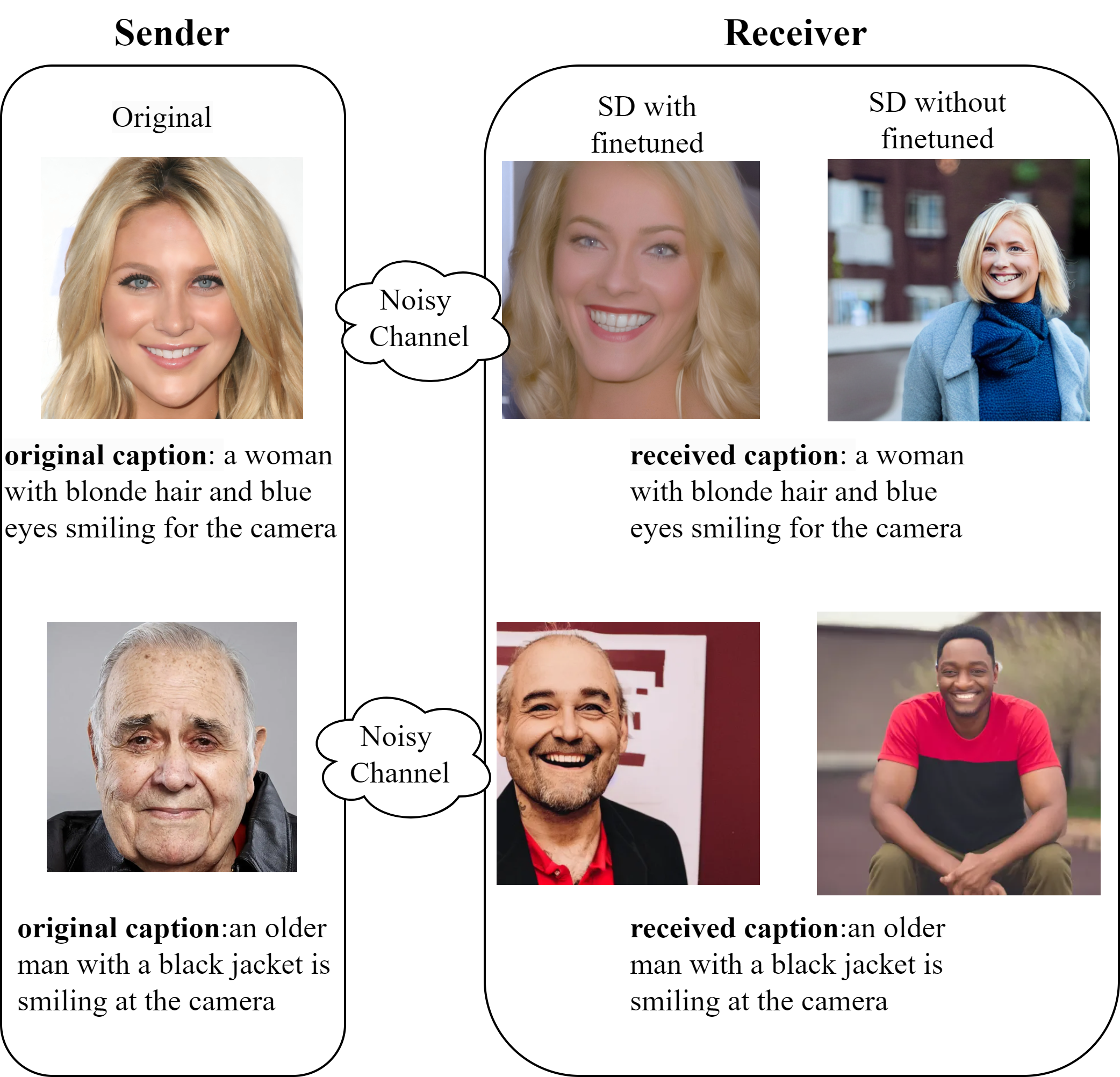}
\centering
\caption{Visual results. On the left side two randomly selected samples along with text which extracted by Img2Txt model. The generated images are on the right. The SNR is 9 dB.}
\label{fig2}
\end{figure}
\vspace{0.4cm}

In a task-oriented communication scenario, SD can struggle to generate specific human portraits or styles without precise control inputs, which makes standard SD less capable of meeting the needs of users, particularly in situations where users have specific requirements for the generated style. Therefore, to produce specific portrait images, we enhanced the model by fine-tuning SD with additional sample data.

We use dreambooth \cite{ruiz2023dreambooth} to fine-tune SD on a portrait dataset \cite{liu2015faceattributes}. We follow the same loss objective as LoRA \cite{hu2021lora}. Let $\Theta(\cdot)$ denote all parameters of a model, $G_s$ denotes the generative model after $s$ iterations, then the hypothesis class at iteration $s$ is:
\begin{equation}
\begin{aligned}
\mathcal{G}_s=\left\{G \mid \operatorname{rank}\left(\Theta(G)-\Theta\left(G_s\right)\right) \leq R\right\},
\end{aligned}
\end{equation}
where $R$ denotes the rank of weight updates and in practice we choose $R = 128$ to balance efficiency and image quality.
This fine-tuning process refines the model's understanding of facial features, expressions, and textures, allowing it to produce images that not only resemble portraits more closely but also convey the subtle qualities and realism that are essential in high-quality portraiture.

\subsection{Semantic Evaluation Metrics}
For text transmission, we use the bilingual evaluation understudy (BLEU) \cite{belghazi2018mutual} score, which is usually used to evaluate the quality of text produced by machine translation systems. It measures how closely the generated text matches reference text by comparing overlapping n-grams in machine translation, which can be described as:
\begin{equation}
    \begin{aligned}
     \log \mathrm{BLEU}=\min \left(1-\frac{l_{\hat{\mathbf{s}}}}{l_{\mathrm{s}}}, 0\right)+\sum_{n=1}^N u_n \log p_n,
    \end{aligned}
\end{equation}
where $u_n$ is the weight of  $n$-grams and $p_n$ is the  $n$-grams score, which is:
\begin{equation}
    \begin{aligned}
     p_n=\frac{\sum_k \min \left(C_k(\hat{\mathbf{s}}), C_k(\mathbf{s})\right)}{\sum_k \min \left(C_k(\hat{\mathbf{s}})\right)},
    \end{aligned}
\end{equation}
where $C_k(\cdot)$ is the frequency count function for the $k$-th elements in $n$-th grams.

For image generation, semantic communication prioritizes semantic-level fidelity over pixel-level fidelity. We evaluate it based on both the quality of the reconstruction and the accuracy of the semantics.
\begin{itemize}
    \item Quality of image reconstruction: We use learned perceptual image patch similarity (LPIPS) \cite{zhang2018unreasonable} metric. which assess the perceptual similarity of intended and generated images using additional neural networks, to evaluate the quality of semantically decoded images.
\end{itemize}
\begin{itemize}
    \item Accuracy of semantics: The accuracy of classifications for expressions, gender, and age can be employed as key indicators when it comes to evaluating the accuracy of semantic reconstruction in portrait image transmission. Therefore, we measure the accuracy of key semantic information reconstruction (including age, gender, and expression) using several neural network classification models.
\end{itemize}

\section{Simulation And Performance Analysis}
We used the BLIP model to generate textual descriptions of 3000 images from the CelebA \cite{liu2015faceattributes} dataset with 3000 sentences. 
Then, we train the BART-SC model on these 3000 sentences with fading channel with $h = 0.9$ and SNR from 5 to 10~dB. 
We employed the Stable Diffusion XL (SDXL) model \cite{podell2023sdxl} as the base model with pre-trained weights and we fine-tuned it with Dreambooth to generate high-quality portrait images. 
The fine-tuning is performed on the default hyperparameters, with a learning rate of 1e-5, and a maximum training epoch deployment of 500. We fine-tuned the SDXL for 10 epochs using 10 images from the CelebA dataset.
For comparison, we evaluate the performance of the proposed Gen-SC by comparing it with standard SD without fine-tuned and traditional text communication link using Huffman coding, (5,7) Reed-Solomon (RS) coding, and 64 quadrature amplitude modulation (QAM), respectively.
\begin{figure}[t]
\centering
\setlength{\belowcaptionskip}{-0.45cm}
\includegraphics[width=8cm]{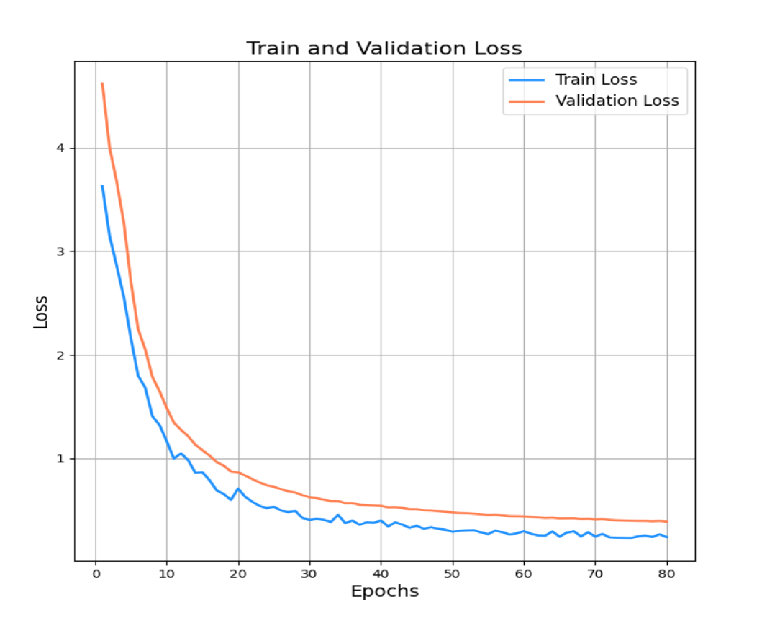}
\centering
\caption{Loss values vs. the number of training epochs}
\label{fig2}
\end{figure}

\begin{figure}[t]
\centering
\setlength{\belowcaptionskip}{-0.45cm}
\includegraphics[width=8cm]{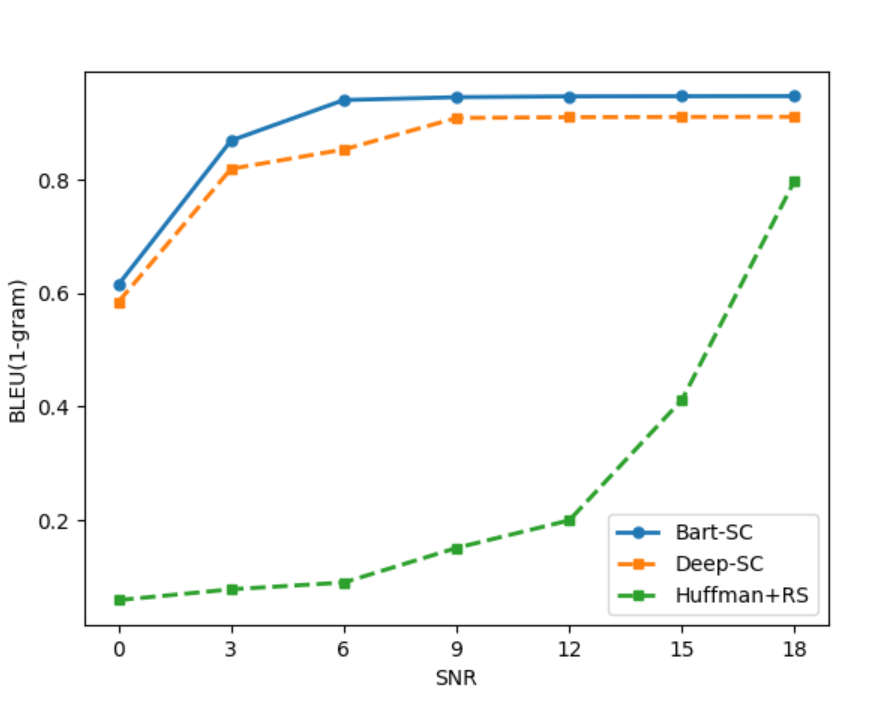}
\centering
\caption{BLEU score vs. SNR for the same total number of transmitted symbols, with Huffman coding with RS (30,42) in 64-QAM, DeepSC and our method.}
\label{fig2}
\end{figure}

In Fig. 6, we showed that the training loss and validation loss change as the number of training epochs varies. Training loss and validation loss decrease as epochs increase. This is due to the fact that the model calculates the loss function to measure the error between the predicted results and the actual labels during training. As the number of training epochs increases, the model parameters are progressively optimized.

In Fig. 7, we showed the relation between the BLEU score and SNR under the same number of transmitted symbols over the AWGN channel.
In the meanwhile, the traditional method uses 64-QAM modulation. 
All BLEU increases as SNR increases. This is due to the fact that when SNR increases, the distortion from noisy channels decreases. 
The BLEU of huffman+RS has the lowest in lower SNR than Bart-SC and Deep-SC. 
This is due to the fact that traditional methods are sensitive to channel error rates caused by channel variations while BART-SC and Deep-SC adopt JSCC that integrates source coding and channel coding into a single model through deep neural networks training, enabling better adaptation to the channel conditions. BART-SC outperforms Deep-SC in each SNR. 
This is due to the fact that BART model learns how to recover the original sequence from a noisy sequence during pre-training, which can enhance the model's robustness to noisy channels.

In Fig. 8, we show how the LPIPS changes as the SNR varies. We observed that the average LPIPS decreases as the SNR increases. Compared to traditional text transmission methods, Gen-SC contributes to a reduction of up to 0.1 in average LPIPS under low SNR conditions, with this reduction diminishing as the SNR increases. This indicates the potential benefits of optimizing Gen-SC levels based on given channel conditions for future research. Additionally, the fine-tuned stable diffusion model performs better in reconstructing images compared to the non-fine-tuned model, as shown in Fig.5.

\begin{figure}[t]
\centering
\setlength{\belowcaptionskip}{-0.45cm}
\includegraphics[width=8cm]{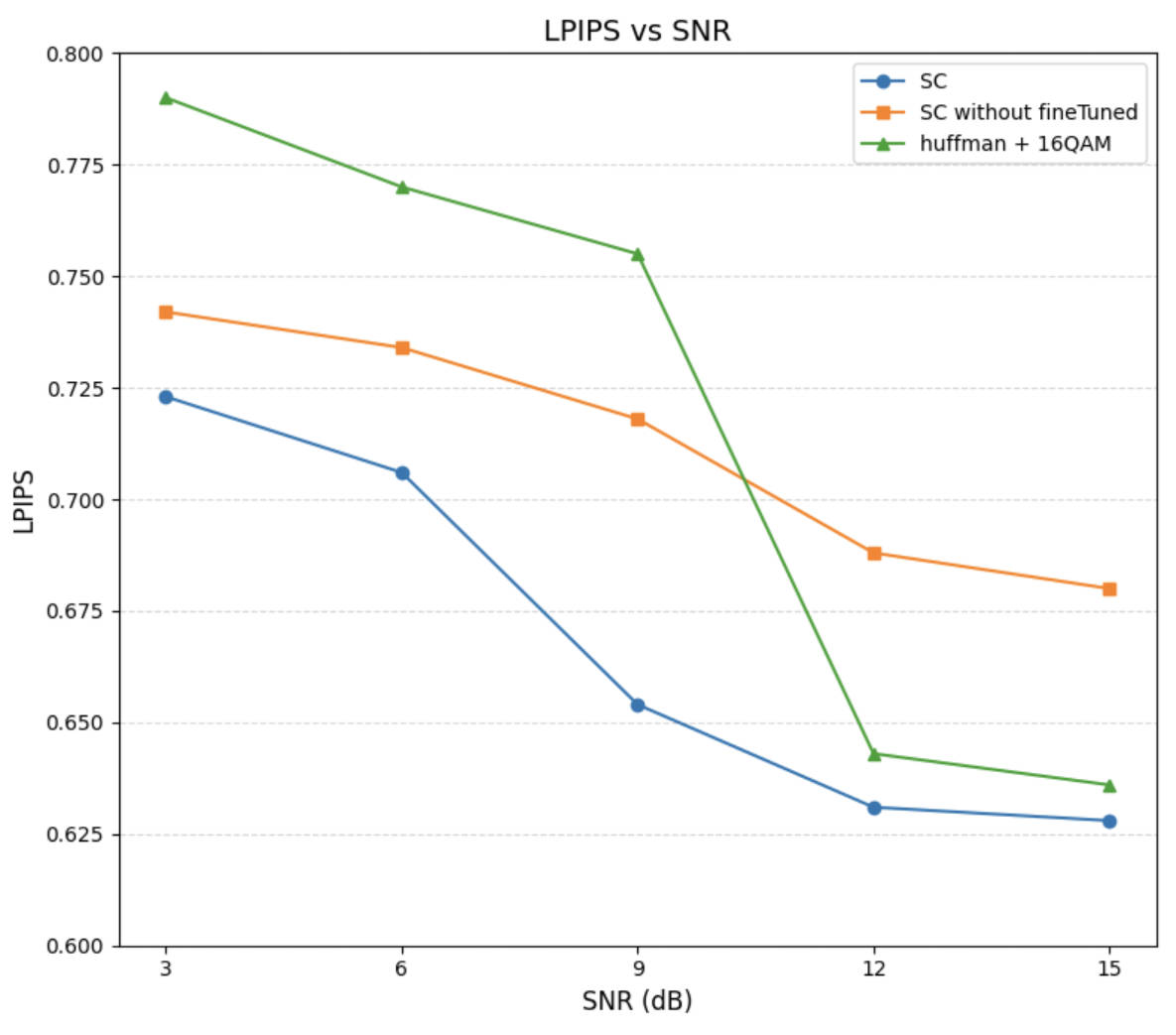}
\centering
\caption{LPIPS vs. SNR.}
\label{fig2}
\end{figure}

In Fig. 9, we show the accuracy of age, gender, and expression in reconstructed images changes as the SNR varies.
We observed that even in low SNRs, the accuracy of all three classifications remains above 80\%.
This reflects the effectiveness of Gen-SC in communication over noisy channels. 
Among them, gender classification achieved the highest accuracy while expression classification had the lowest accuracy. 
This is due to the fact that SD tends to reconstruct some neutral expressions as negative expressions. 
For instance, a "serious look" might be reconstructed as an image with a frowning expression. 
The accuracy of age reconstruction was intermediate, with the testing revealing that the granularity for age reconstruction is relatively coarse.

\begin{figure}[t]
\centering
\setlength{\belowcaptionskip}{0.0cm}
\includegraphics[width=8cm]{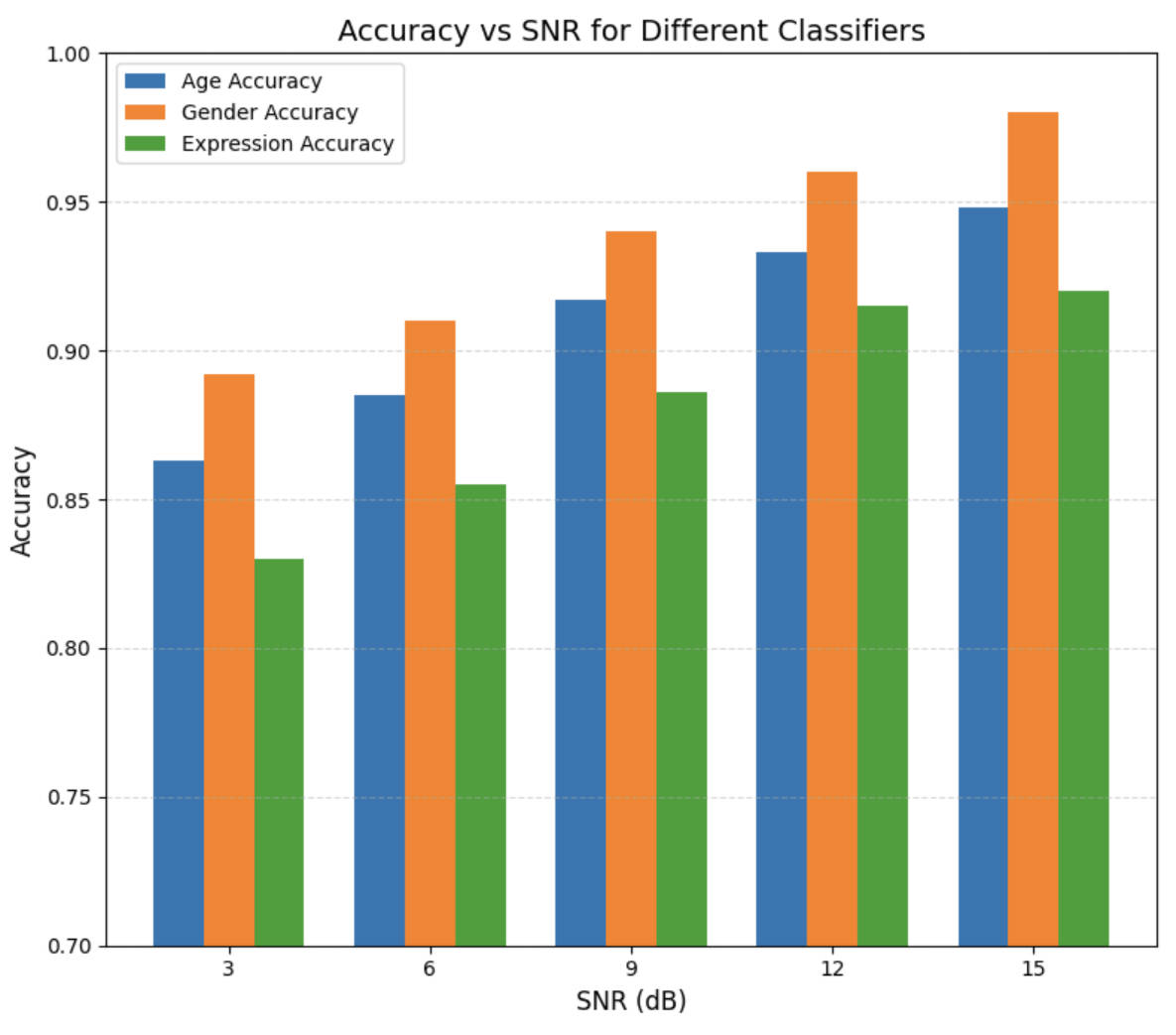}
\centering
\caption{Accuracy of age, gender and expression in reconstructed images versus the SNR}
\label{fig2}
\end{figure}

\section{Conclusion}
In this paper, we proposed a Gen-SC framework for scenarios involving portrait transmission to achieve an efficient and robust semantic communication system. 
The main processes in Gen-SC include converting images into text, utilizing a transformer-based text transmission model, and employing a diffusion model for image reconstruction. 
Experimental results indicate that the proposed framework can significantly reduce the amount of transmitted data while preserving the semantic information.
Additionally, the transformer-based transmission model provides better robustness to noisy wireless channels during text transmission than baseline methods.
Finally, the scheme of fine-tuning the diffusion model enhanced the perceptional similarity in portrait image generation.


\bibliographystyle{IEEEtran}
\renewcommand{\baselinestretch}{1.38}
\bibliography{BNN}
\end{document}